# Structural and electronic properties of epitaxial YBa$_2$Cu$_3$O$_{7-\delta}$ –La$_{0.67}$Ca$_{0.33}$MnO$_3$ bilayers grown on SrTiO$_3$ (110) substrates

L. Mustafa[1], N. Driza[1], S. Soltan[1,2], M. Le Tacon[1], H.-U. Habermeier[1], B. Keimer[1]

[1]Max-Planck-Institute for Solid State Research, Heisenbergstr. 1, D-70569 Stuttgart, Germany
[2]Faculty of Science, Helwan University, 11795-Cairo, Egypt

## Abstract

Epitaxial bilayers of the high-temperature-superconductor YBa$_2$Cu$_3$O$_{7-\delta}$ (YBCO) and the ferromagnetic metal La$_{0.67}$Ca$_{0.33}$MnO$_3$ (LCMO) were prepared by pulsed laser deposition on (110)-oriented SrTiO$_3$ substrates, such that the CuO$_2$ planes of YBCO are perpendicular to the YBCO-LCMO interface. X-ray diffraction and Raman scattering demonstrate complete (110) orientation of both YBCO and LCMO overlayers. The resistivity and magnetization of the bilayer films are highly anisotropic. The critical temperatures for superconductivity and ferromagnetism as well as the saturation magnetization exhibit modest reductions compared to corresponding bulk values.

Superlattices and heterostructures of transition metal oxides with strong electron correlations are gaining increasing interest as a platform to harness a variety of quantum many-body phenomena (including metal-insulator transitions, ferromagnetism, superconductivity, multiferroicity, as well as charge, spin, and orbital ordering) for potential applications.[1,2] In structures composed of two or more TMOs, interference between these states across interfaces has the potential to generate new phenomena and functionalities. In particular, hybrid structures of high-temperature superconducting $YBa_2Cu_3O_{7-\delta}$ (YBCO) and half-metallic ferromagnetic $La_{0.67}Ca_{0.33}MnO_3$ (LCMO) have been used to explore possible interference between these antagonistic order parameters.[4-16] Research on such structures has revealed charge transfer across the YBCO-LCMO interfaces[6-8] as well as interfacial spin[9-12] and orbital[7] polarization, with consequences for their macroscopic properties including charge transport[4,6] and thermoelectricity.[13] Hybridization of phonon modes[14] and electronic interference phenomena[15] over a range of several tens of nanometers have also been observed. However, most of the superlattices and heterostructures thus far investigated were grown along the (001) axis of the YBCO crystal structure, such that the direction perpendicular to the highly conducting $CuO_2$ planes is parallel to the YBCO-LCMO interface. Proximity coupling of superconducting and ferromagnetic order parameters has not yet been directly identified in these structures,[16] presumably as a consequence of the small superconducting coherence length $\xi_c \sim 0.3$ nm of YBCO in the [001] direction,[17] in combination with the depleted density of mobile charge carriers and the spin and orbital polarization close to the interface.

In view of the much larger coherence length $\xi_{ab} \sim 1.6$ nm along the $CuO_2$ planes of YBCO,[17] prospects for proximity effects are far brighter for YBCO-LCMO hybrids grown in such a way that these planes are perpendicular to the interfaces. With this motivation, we have developed a procedure for the growth of YBCO-LCMO bilayers on (110)-oriented $SrTiO_3$ (STO) substrates. Whereas several groups have succeeded in growing single-layer YBCO films on (110)-STO substrates using either the $PrBa_2Cu_3O_{7-x}$ (PBCO) hetero-templating or the YBCO self-templating technique, [18-20] only a few reports dealing with (110)-oriented cuprate-manganate bilayers [21,22] or trilayers[23] can be found in the literature. Mandal and coworkers[21] used pulsed laser deposition to grow (110)-oriented YBCO (100nm) -LSMO (30nm) heterostructures on (110)-STO substrates with a 20 nm PBCO template. However, X-ray diffraction showed that more than a third of the film volume comprised (103)- or ($10\bar{3}$)-oriented YBCO grains. Measurements of the electronic properties of such mixed-phase structures are difficult to interpret. Tse *et al.*[22] characterized YBCO-LCMO bilayers deposited on PBCO-buffered (110)-STO by electron microscopy, but measurements of the macroscopic phase composition and electronic properties of these structure have not been reported.

In this Letter, we describe the growth of single-phase (110)-oriented YBCO-LCMO bilayers deposited on PBCO-buffered (110)-oriented STO substrates. The films were deposited using pulsed laser deposition with parameters optimized for the growth of (110)-oriented YBCO films and standard parameters for the growth of LCMO films. Specifically, a 30 nm thick PBCO template was grown at 650 °C at a rate of 0.06 nm/sec, using a photon fluency of 1.5 J/cm$^2$ in an oxygen pressure of 0.4 mbar. YBCO layers of varying thickness were then deposited at 730 °C at a rate of 0.06 nm/sec. Finally, LCMO layers were deposited at a rate of 0.048 nm/sec at the same temperature. After deposition, the samples were cooled to 530 °C in 0.4 mbar oxygen at a rate of 20 °C/min, annealed in 1 bar oxygen for one hour, and then cooled to room temperature.

The inset of Fig. 1 shows a sketch of the resulting structure for a YBCO layer thickness of 100 nm and an LCMO layer thickness of 50 nm. The X-ray diffraction pattern displayed in the main panel of Fig. 1 exhibits sharp, intense doublets at scattering angles 2θ ~32.5° and ~ 69°. The higher-intensity component comes from the (110) and (220) planes of the substrate, whereas the lower-intensity peaks arises from the LCMO/YBCO/PBCO film stack and can be assigned to the (110) and (220) reflections, respectively. These data confirm the absence of YBCO (001) grains as well as materials purity (i.e. the absence the absence of mixed composites). Heterostructures with LCMO layer thicknesses between 50 and 100 nm and YBCO layer thicknesses between 30 and 100 nm showed similar results.

Since the (220) reflection of LCMO and the (110) and (103) reflections of YBCO appear at nearly identical 2θ-values in X-ray diffraction, additional measurements are required to confirm the (110) phase purity. To this end, pole figures in the YBCO (117) orientation were chosen, because there is no overlap with any of the STO or LCMO reflections (Fig. 2a). On the other hand, reflections arising from (103)- or ($10\bar{3}$)-oriented grains of YBCO would be clearly visible in addition to the YBCO (110) peaks The absence of such reflections in the pole figures clearly demonstrates that the films are exclusively (110) oriented, within the detection limit of x-ray diffractometry.

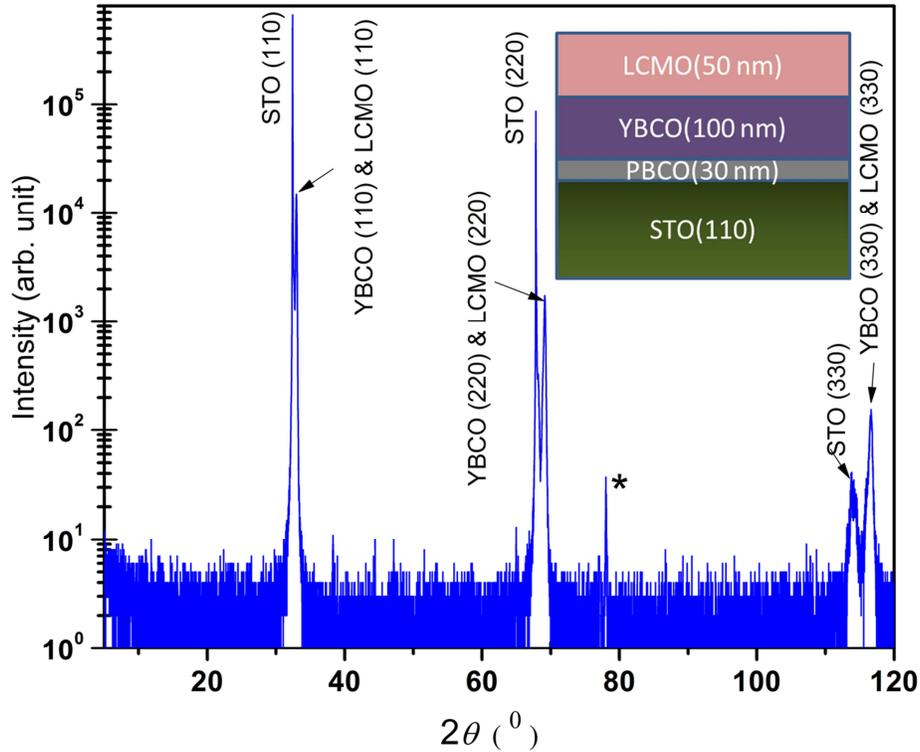

Fig.1. θ - 2θ x-ray diffraction pattern of an LCMO (50 nm) – YBCO (100 nm) heterostructure grown on (110)-STO (see sketch in the inset).The peak marked with the asterisk is due to the sample holder of the diffractometer.

To check the orientation of the LCMO layer with respect to the substrate and the YBCO layer, the (310) plane of LCMO was chosen for an additional set of pole figures (Fig. 2b). Here, 12 peaks corresponding to (310) and equivalent planes of LCMO are observed; note that these overlap with those of the STO substrate. No other peaks of LCMO are observed, indicating that the LCMO layer is grown epitaxially with its (110) plane parallel to the (110) plane of STO. The pole figure analysis of the heterostructures thus demonstrates that the (110) planes of the STO substrate, the YBCO layer, and the LCMO layer are parallel to each other.

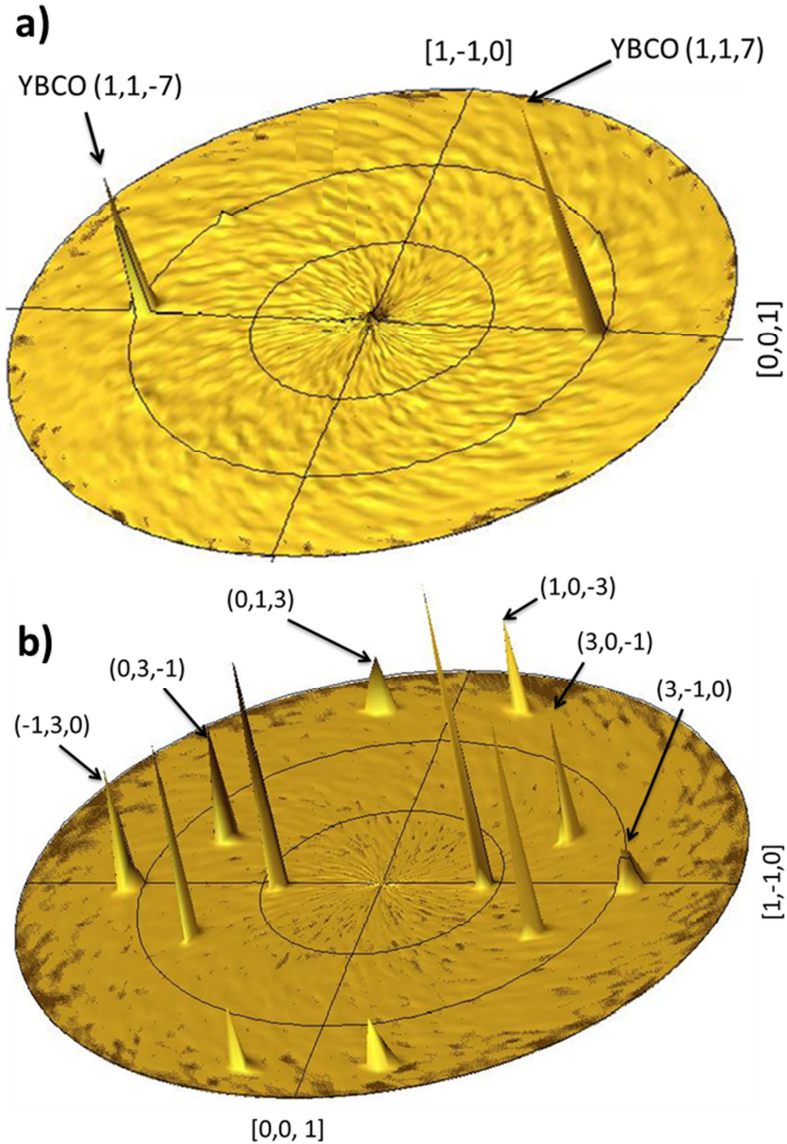

Fig. 2. X-ray pole figures of an LCMO (50 nm) – YBCO (100 nm) heterostructure (a) around the (117) reflection of YBCO, and (b) around the (103) reflection of LCMO. Note that the reflections in panel (b) overlap with those of the STO substrate.

The oxygenation state and orientational perfection of the YBCO layer were further investigated by Raman spectroscopy. Figure 3 shows Raman spectra with photon polarization parallel (*zz* configuration) and perpendicular (*xx/yy* configuration) to the YBCO *c*-axis. The frequency of the apical oxygen vibration around ~500 cm$^{-1}$, which is very sensitive to doping,[24] indicates a slight oxygen deficiency of the YBCO layer in our film ($\delta \sim 0.1$). The planar oxygen vibration of YBCO at ~340 cm$^{-1}$ is absent (within the experimental sensitivity) in *zz* geometry, and the intensity of the planar Cu vibration at ~ 150 cm$^{-1}$ is greatly reduced in *xx/yy* geometry, in accordance with the established Raman selection rules.[25,26] This confirms the (110) orientation of the YBCO layer, in agreement with the x-ray analysis.

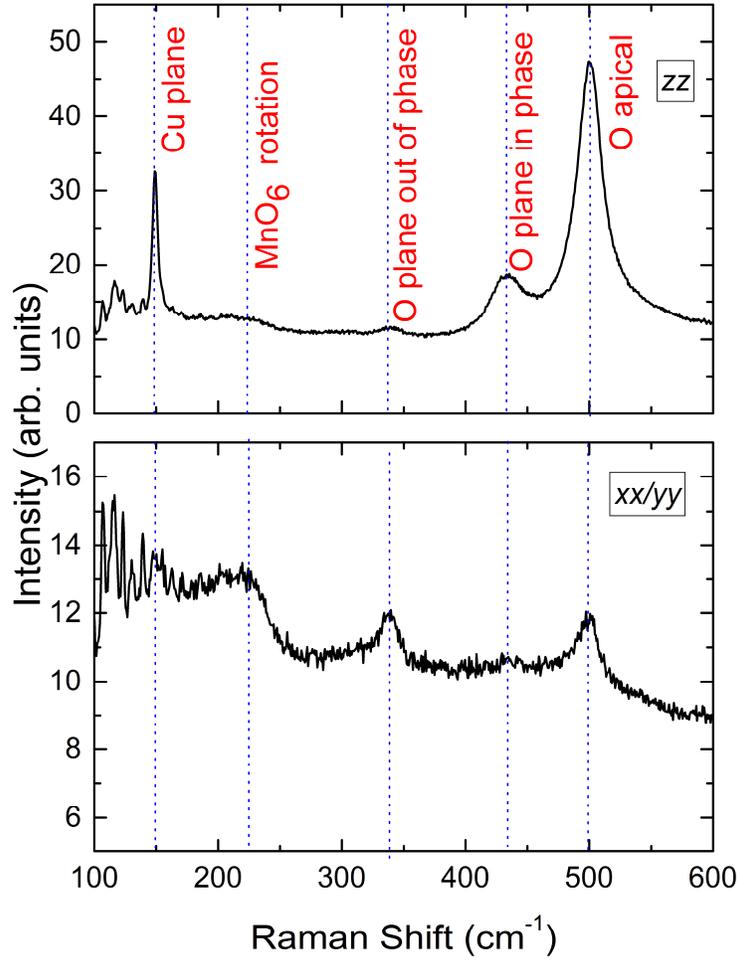

Fig. 3: Polarized Raman scattering spectra of an LCMO (50 nm) – YBCO (100 nm) heterostructure in a scattering geometry with incoming and reflected light parallel (top) and perpendicular (bottom) to the YBCO *c*-axis. The data were taken with a photon wavelength of 514.52 nm. The weak intensity at 340 cm$^{-1}$ in *zz* geometry is due to polarization leakage. The low-energy intensity seen in *xx/yy* geometry is due to scattering from air.

YBCO films deposited onto PBCO buffered (110)-oriented STO substrates as well as our bilayers show a relatively poor surface roughness of ~ 5 nm. In order to check the YBCO-LCMO interface quality we performed soft x-ray resonant reflectometry measurements (using a photon energy corresponding to the Mn L3-edge) in (110) -oriented YBCO-LCMO-YBCO trilayers prepared under the same conditions as the bilayer films reported here. They show roughness values of ~ 3nm for the whole film area of 5x5 mm$^2$. TEM analysis of these films indicate at a local scale ( >70 nm ) an interface roughness of ~ 1nm.[27]

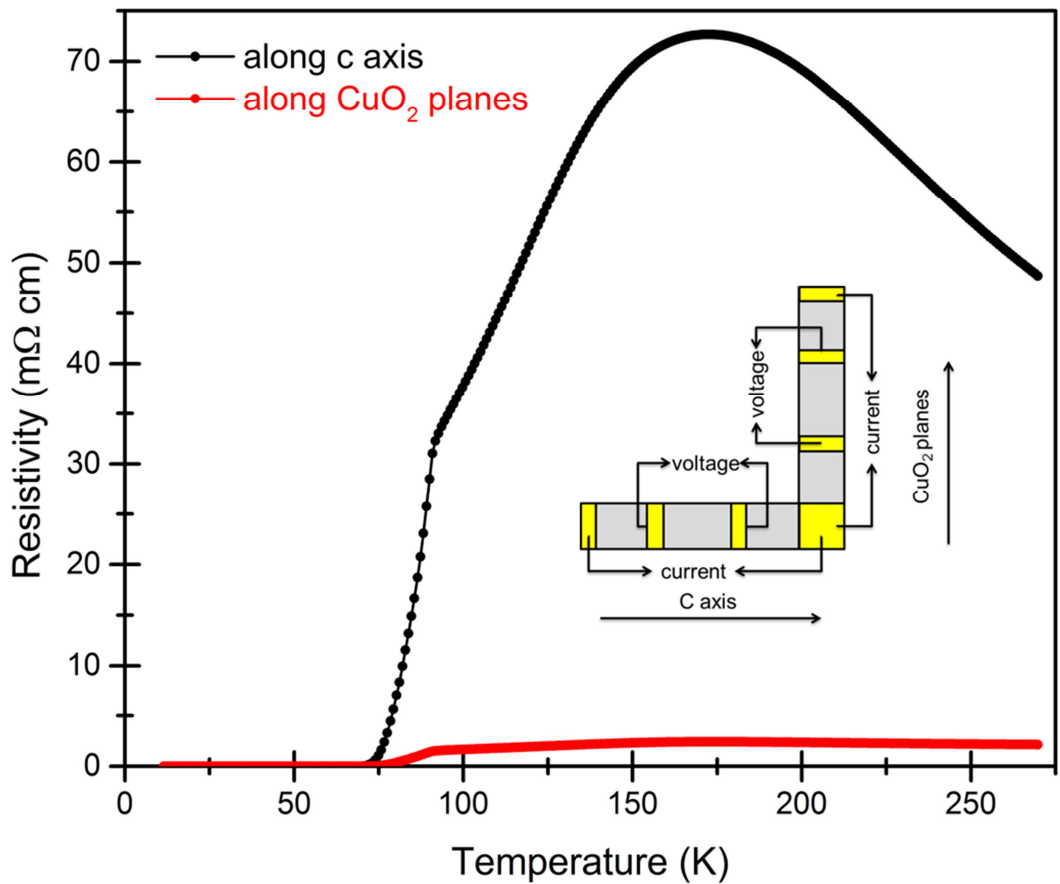

Fig. 4: Temperature dependence of the resistivity of an LCMO (50 nm) – YBCO (100 nm) heterostructure parallel and perpendicular to the YBCO $c$-axis. The inset shows the arrangement of Au contact pads evaporated on the film surface.

Figure 4 presents the temperature dependence of the resistivity measured perpendicular ($\rho_{ab}$) and parallel ($\rho_c$) to the YBCO c-axis using a standard four-probe arrangement (inset in Fig. 4). The large anisotropy ($\rho_c/\rho_{ab}$ ~ 25 at room temperature, ~ 35 at 180 K) is consistent with the single-phase nature of the film inferred from the structural analysis and is caused by the resistivity anisotropy of YBCO. The temperature dependence of the resistivity of the film can be modeled as a network of parallel resistors and is hence dominated by the layer with the smallest resistivity. In the direction of the $CuO_2$ planes the value is dominated by the YBCO layer whereas in the perpendicular direction both resistivities ( YBCO and LCMO, respectively) are of comparable magnitude and contribute nearly equally to the total value. The contribution of the LCMO layer can be recognized in the resistivity peak around the Curie-temperature that is typical for ferromagnetic LCMO films. This contribution is presumably responsible for the lower resistivity anisotropy of the film compared to bulk YBCO, where $\rho_c/\rho_{ab}$ ~ 100 at optimal doping. Both resistivity curves reveal a superconducting transition with an onset around 90 K, and zero resistivity (within the measurement error) around 75 K. The rounding of the superconducting transition may reflect intrinsic interfacial effects (such as the transfer of spin-polarized quasiparticles across the interface[28]) and/or residual inhomogeneity of the oxygen content within the YBCO layer.

Figure 5 shows the magnetic moment of the film measured in a magnetic field H= 10 Oe applied in different directions. The superconducting transition is most readily apparent if the field is applied perpendicular to the film plane (main panel of Fig. 5). The critical temperature determined in this way coincides with the zero-resistance state determined by the transport measurements (Fig. 4). When the field is applied in the film plane, the signature of superconductivity in the magnetization data is more subtle, due to the smaller screening current loops for fields applied in this direction.  One the other hand, for this field direction the shape anisotropy that confines the ferromagnetic moment of the LCMO layer to the film plane greatly facilitates the determination of the Curie temperature, which is observed to be ~ 220 K – close to (but somewhat lower than) the bulk value of ~ 270 K. Likewise, the saturation magnetization of our film (~ 2 µB per Mn atom; data not shown) is lower than the corresponding value in bulk crystals and in bulk-like LCMO films.[29] The difference in the magnetization of the film measured along and perpendicular to the CuO2 planes ( see inset in Fig. 5 ) is ascribed to its crystalline anisotropy. The M(T) measurements have been performed at an applied field of 10

Oe, i.e. much lower than the coercive field of the film, consequently their saturation values are different. The measurements indicate that the magnetic easy axis is aligned to the [1-10] direction of the LCMO layer in accordance with the results given by Infante et al.[30]

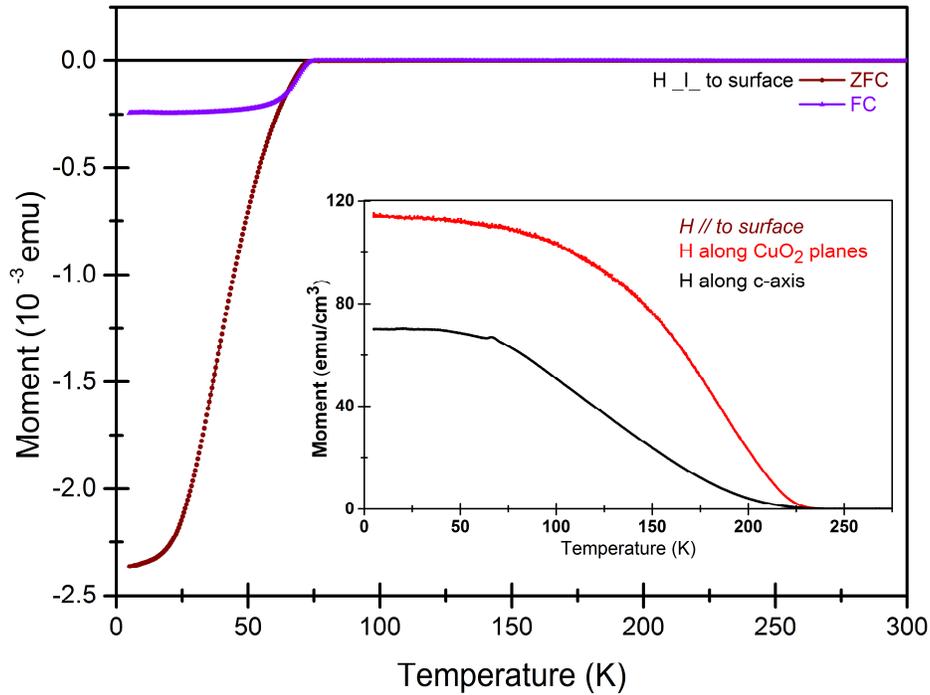

Fig.5: Temperature dependence of the magnetic moment of an LCMO (50 nm) – YBCO (100 nm) heterostructure determined by Superconducting Quantum Interference Device (SQUID) magnetometry in a magnetic field H=10 Oe applied perpendicular to the film plane. The inset shows a measurement with H=10 Oe applied parallel to the film plane, either parallel or perpendicular to the YBCO *c*-axis.

Further work is required to assess whether the reduction of the Curie temperature and saturated moment has an intrinsic origin, or whether it is due to a slight off-stoichiometry of the LCMO layer in the film.

In summary, we have shown that our YBCO-LCMO bilayers grow epitaxially, with the (110) planes of LCMO, YBCO, and STO parallel to each other. Admixtures of (001) and/or (103) grains of YBCO are below our detection limit. The charge transport and magnetic properties of the film are highly anisotropic, as expected based on the anisotropic electronic structure of YBCO. The deposition procedure we established thus opens new perspectives for the exploration of the interplay between high-temperature superconductivity and ferromagnetism in metal-oxide heterostructures.

The authors thank G. Logvenov for valuable discussions, and G. Christiani for technical support. Financial support was provided by the German Science Foundation under collaborative research Grant No. SFB/TRR 80.